# Quantifying the informativeness for biomedical literature summarization: An itemset mining method

Milad Moradi, Nasser Ghadiri[1]

*Department of Electrical and Computer Engineering, Isfahan University of Technology,*

*Isfahan 84156-83111, Iran*

E-mail: milad.moradi@ec.iut.ac.ir, nghadiri@cc.iut.ac.ir

*Abstract*

*Objective*: Automatic text summarization tools can help users in the biomedical domain to access information efficiently from a large volume of scientific literature and other sources of text documents. In this paper, we propose a summarization method that combines itemset mining and domain knowledge to construct a concept-based model and to extract the main subtopics from an input document. Our summarizer quantifies the informativeness of each sentence using the support values of itemsets appearing in the sentence.

*Methods*: To address the concept-level analysis of text, our method initially maps the original document to biomedical concepts using the Unified Medical Language System (UMLS). Then, it discovers the essential subtopics of the text using a data mining technique, namely itemset mining, and constructs the summarization model. The employed itemset mining algorithm extracts a set of frequent itemsets containing correlated and recurrent concepts of the input document. The summarizer selects the most related and informative sentences and generates the final summary.

---

[1] *Corresponding author. Address: Department of Electrical and Computer Engineering, Isfahan University of Technology, Isfahan 84156-83111, Iran. Phone : +98-31-3391-9058, Fax: +98-31-3391-2450, Alternate email: nghadiri@gmail.com*

*Results*: We evaluate the performance of our itemset-based summarizer using the Recall-Oriented Understudy for Gisting Evaluation (ROUGE) metrics, performing a set of experiments. We compare the proposed method with GraphSum, TexLexAn, SweSum, SUMMA, AutoSummarize, the term-based version of the itemset-based summarizer, and two baselines. The results show that the itemset-based summarizer performs better than the compared methods. The itemset-based summarizer achieves the best scores for all the assessed ROUGE metrics (R-1: 0.7583, R-2: 0.3381, R-W-1.2: 0.0934, and R-SU4: 0.3889). We also perform a set of preliminary experiments to specify the best value for the minimum support threshold used in the itemset mining algorithm. The results demonstrate that the value of this threshold directly affects the accuracy of the summarization model, such that a significant decrease can be observed in the performance of summarization due to assigning extreme thresholds.

*Conclusion*: Compared to the statistical, similarity, and word frequency methods, the proposed method demonstrates that the summarization model obtained from the concept extraction and itemset mining provides the summarizer with an effective metric for measuring the informative content of sentences. This can lead to an improvement in the performance of biomedical literature summarization.

*Keywords*: Biomedical text mining; Data mining; Frequent itemset mining; Informativeness; Concept-based text analysis; Domain knowledge;

## 1. Introduction

With the fast growth in producing digital text content in the biomedical domain, much effort has been made toward developing automatic tools to alleviate the difficulties of finding relevant information. Notably, clinicians and researchers are often faced with a common problem, i.e. acquiring the intended information from the vast sources of available text documents. The information is obtained from various resources, such as scientific literature databases, clinical trials, Electronic Health Record (EHR) systems, multimedia documents, e-mailed reports, and web documents [1, 2]. Review of the biomedical literature is an essential prerequisite to conducting the main steps of research and experiment, such as getting to know the state-of-the-art, collecting the required information, making new hypotheses, and interpreting the results of experiments [3]. However, identifying useful information within the large volume of biomedical literature is an exhausting and time-consuming process. Automatic text summarization makes the task easier by condensing the source text while preserving the content that refers to the essential points of text. Reeve et al. [4] have cited five reasons for generating summaries of full-text papers even with the presence of the abstracts: (1) There are variants of an ideal summary in addition to the abstract; (2) Some content of the full-text may be missed in the abstract; (3) Question answering systems can make use of customized summaries to provide personalized information; (4) Automatic summaries allow the abstract services to scale the number of documents they can evaluate; and (5) Investigating the quality of sentence selection methods can be helpful in the development of multi-document summarization systems [4].

In a typical summarization system, the source text is first analyzed and modeled by a particular method; then the final summary is generated. The majority of summarization systems deal with the source text through constructing a model from the words. However, in the biomedical domain, it seems that the term-level analysis of text may not yield adequate summarization performance regarding informative content conveyed by summaries [4, 5]. The biomedical literature (similar to the scientific literature of many other fields) has some properties, such as the variety of synonyms and homonyms, elisions, and abbreviations [5]. More specifically, text documents related to each subdomain of biomedical sciences have their particular singularities that should be taken into account. For example, in the genome sciences, each distinct gene may have multiple synonyms [6]. When a text document from the genome domain is analyzed at term-level, multiple synonymous names of a unique gene may be considered as different entities, and their relations are overlooked. On the other hand, in the concept-level analysis of text, multiple terms that share the same meaning are considered as a single entity and will be referred to as a single concept. Therefore, the accuracy of the model improves, enhancing the quality of summarization [4, 5, 7, 8]. To address the properties of biomedical literature, the concept-level analysis of text is performed exploiting domain-specific biomedical knowledge sources, such as ontologies, vocabularies, and taxonomies.

A summarization system decides which sentences are relevant and informative, and selects them to be included in the summary according to the model produced from the source text. Using an appropriate metric for measuring the informativeness of sentences, a summarization system can generate a useful summary that covers the main points of the original text. Several summarization methods measure the quality of sentences based on some statistical and general metrics. Typical metrics include the position of sentences, the frequency of words, the presence of positive and negative keywords, the length of sentences, the similarity of sentences to the document title, the existence of numerical data, and the presence of proper nouns [9-11]. Although these methods have achieved a desirable summarization quality compared with their counterparts, they may not suit the specific requirements of text modeling and sentence selection in biomedical summarization. We show that it is required to construct an accurate concept-based model from the source text so that it can be effectively used to measure the relatedness and informativeness of sentences in this type of summarization.

In this paper, we propose a novel biomedical text summarization method employing the concept-level analysis of text together with a data mining approach, namely itemset mining. The goal of our proposed itemset-based summarizer is to generate an accurate concept-based model from the source text. The produced model represents the main subtopics of text and a measure of their importance in the form of extracted frequent itemsets. The itemset-based summarizer, first, maps the input text document to biomedical concepts contained in the Unified Medical Language System (UMLS) [12], a popular knowledge source in the biomedical domain. Then, it generates a transactional data representation from the source text and the extracted concepts. Afterward, it discovers the main subtopics of the document using a well-established data mining technique, namely itemset mining [13]. To this aim, we utilize the Apriori algorithm [14]. Each extracted frequent itemset is a set of correlated concepts that frequently occur in the source text, and the summarizer uses them to quantify the informativeness of the sentences. Eventually, the itemset-based summarizer assigns a score to each sentence. The assigned scores measure the informative content of the sentences. The summarizer selects the most relevant and informative sentences and puts them together to form the final summary.

To evaluate the performance of the proposed itemset-based summarizer, we perform a set of experiments on a corpus of biomedical scientific papers. The experimental results show that our proposed biomedical summarization method performs better than the statistical and similarity feature-based, word frequency, and baseline methods regarding the Recall-Oriented Understudy for Gisting Evaluation (ROUGE) metrics [15].

The remainder of this paper is organized as follows. A background of text summarization and previous works in biomedical summarization is given in Section 2. In Section 3, we explain our proposed biomedical text summarization method, as well as the evaluation methodology. The results of experiments and assessments are presented in Section 4 and are discussed In Section 5. Finally, we draw some conclusions and point out the potential future work in Section 6.

## 2. Background

First research efforts in automatic text summarization were commenced in the late 1950s and 1960s by the early works of Luhn [16] and Edmundson [17]. During the last two decades, various summarization methods have been proposed using statistical, Natural Language Processing (NLP), machine learning, graph-theoretic, artificial intelligence, and mathematical approaches [18 , 19]. Automatic text summarization systems are designed to generate a short form of text that conveys the main ideas of the original document [18]. Automatic text summarizers are classified into *abstractive* and *extractive* systems. An abstractive summarizer generates a new content that conveys the implications of the source text by employing NLP and linguistic methods. An extractive summarizer produces a shorter version of the input text through selecting the most representative sentences from the original wording and putting them together [20]. Regarding the number of documents that a summarization system receives as input to produce a unified summary, summarization methods are divided into *single-document* and *multi-document* methods [1]. *Generic* versus *query-focused* is another categorization for text summarization. A generic summary gives a general implication of the information provided by the source text, while a query-focused summary provides a summary that contains the content related to a given query [2]. Summaries can be styled as *indicative* or *informative* outputs. Indicative summaries provide information about the topics of the input document pointing to some parts of the text. Users still need to read the original text to acquire sufficient information. Informative summaries provide complete and adequate content about the topics of text so that it is not necessary to retrieve the primary document [18, 19]. Our proposed method is extractive, informative, generic, and single-document.

Several summarization methods, like SweSum [21] and SUMMA [9], employ statistical, term-based similarity, and word frequency features. We show that in biomedical literature summarization, the use of traditional statistical and word frequency features should be replaced by another metric that can adequately measure the extent of informative content of each sentence. Such a parameter can improve the accuracy of summarization, because it evaluates the quality of sentences according to their approximated meaning rather than their position, length, and contained terms. We address this problem by introducing a concept-based summarization model that can be useful for assessing the informative content of sentences in biomedical literature summarization.

Various domain-independent text summarization methods have been proposed using clustering methods [22], genetic algorithms [23], graph-based methods [24], Latent Semantic Analysis (LSA) [25], complex networks [26], optimization methods [27], topic based approaches [28], statistical approaches [29]. However, as noted earlier, the biomedical domain has some singularities that require specialized summarization methods [2, 5]. Domain knowledge resources can be used in text analysis process to build a rich representation of the source text and improve the performance of biomedical summarization. One of the useful sources of knowledge in the biomedical domain is the UMLS [12], developed by the US National Library of Medicine.

The UMLS is a reputable biomedical knowledge source that unifies over 100 controlled vocabularies, classification systems, and additional sources of knowledge. It has been employed by some summarization systems [4, 5, 7, 8, 30-32]. The UMLS contains three main components: the Specialist Lexicon, the Metathesaurus, and the Semantic Network. The *Specialist Lexicon* is considered as a lexicographic information database and includes English and biomedical vocabularies [33]. The *Metathesaurus* is a large, multi-purpose, and multi-lingual thesaurus that includes a large number of biomedical and health related concepts, along with their synonyms and relationships [34]. The *Semantic Network* contains a set of broad subject categories, called semantic types to divide the concepts of the Metathesaurus into stable categorizations [35]. It also defines a set of valuable relationships between the semantic types. Plaza [36] investigated the impact of a set of knowledge sources on the performance of biomedical summarization. She tested different combinations of particular sources in the UMLS for retrieving biomedical domain concepts. The results showed that the quality of produced summaries could be significantly improved by using an appropriate source of knowledge.

In the last two decades, various biomedical text summarization methods have been proposed using different approaches [1, 2]. Some methods have addressed this type of summarization using concept-based methods. Plaza et al. [5] performed biomedical summarization using a semantic graph-based approach. Their summarization method used the concepts and semantic relations of the UMLS to construct a graph. Then, it applied a degree-based clustering algorithm to identify different themes within the document. The authors evaluated three different sentence selection heuristics to investigate the impact of various selection strategies on the quality of summarization. Another concept-based biomedical summarization system [30] made use of genetic clustering approach in combination with graph connectivity information. It employed genetic graph clustering to identify the topics of the document according to its concepts. It also used the connectivity information of the graph to assess the relevance of different subjects. ChainFreq [4] is a hybrid biomedical summarization method that combined BioChain [31] and FreqDist [32] summarizers. ChainFreq method used BioChain to identify thematic sentences. In the hybrid method, FreqDist was responsible for removing information redundancy. Both BioChain and FreqDist methods utilized the UMLS concepts to improve their topic and frequency modeling approaches. Sarkar [37] identified a vocabulary of cue terms and phrases from the biomedical domain and incorporated the vocabulary as a source of domain knowledge in his feature-based summarization method. The method used two new additional features, namely the presence of cue medical terms and the existence of new terms, along with a set of traditional features, such as word frequency, sentence length, sentence position, and the similarity with the title. The summarizer assigned a score to each sentence and selected the high-scoring sentences to form the summary. As reported by the above methods, when an appropriate summarization approach was combined with domain knowledge, the biomedical summarizers could perform better than their domain-independent counterparts. Our proposed method employs the UMLS knowledge source to improve its itemset mining model. The impact of both concept-based and term-based approaches on the performance of the itemset-based summarizer will be discussed in Section 5.

Recently, a few domain-independent summarization methods have addressed the challenges of selecting the most relevant and non-redundant sentences in multi-document and multilingual summarization using frequent itemsets [38-40]. Frequent itemset mining is a data mining technique that can be used to discover frequent patterns in a dataset. It has already been used in summarization of transactional [41] and medical data [42]. However, so far it has not been used for biomedical text summarization. In our proposed biomedical summarization method, we employ frequent itemset mining to extract the subtopics of an input document. In the sentence scoring stage, we use the extracted itemsets to measure the informative content of sentences.

## 3. Methods

In this section, we explain our itemset-based biomedical text summarization method. We also describe the evaluation methodology in this section.

### 3.1. Itemset-based summarizer

The itemset-based summarizer performs the summarization task in four steps, including (1) preprocessing, (2) data preparation for itemset mining (3) extracting main subtopics, and (4) sentence scoring and summary creation. Fig. 1 illustrates the architecture of the itemset-based summarizer. In the following subsections, we give the detailed description of each step.

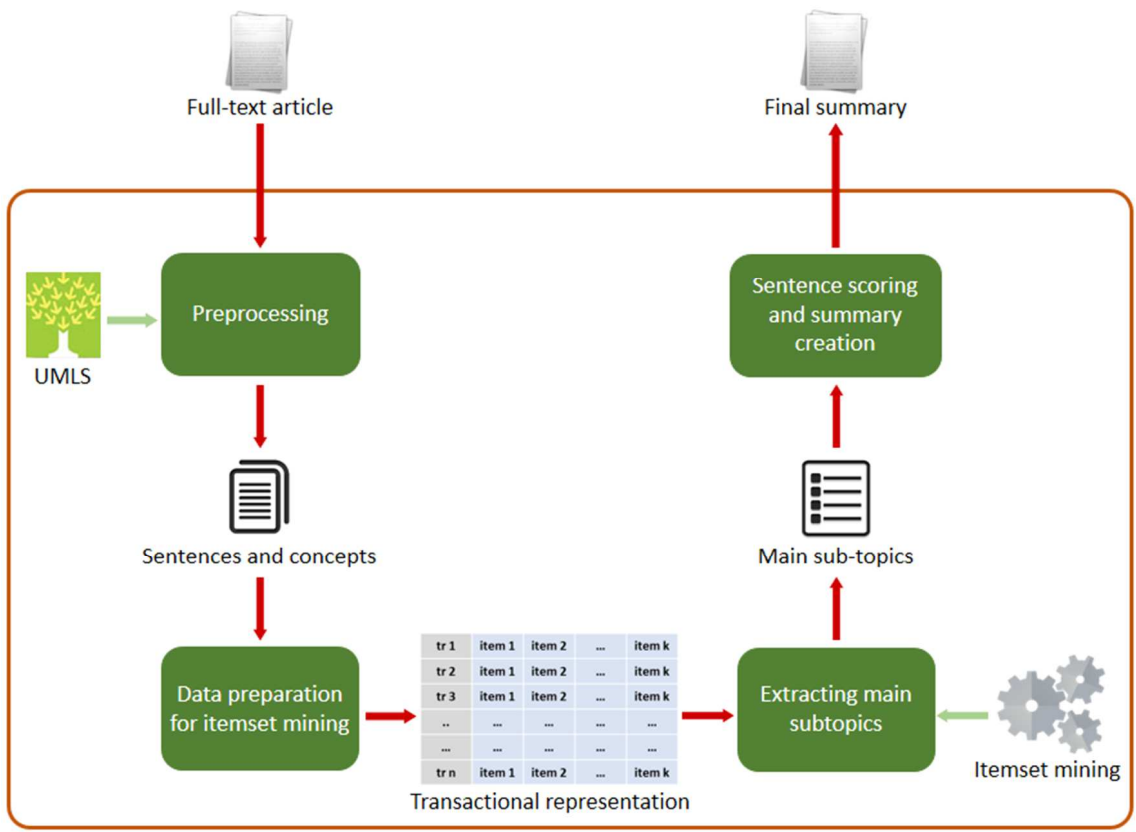

**Fig. 1.** The architecture of the itemset-based biomedical text summarizer.

### 3.1.1. Preprocessing

Before performing the task of summarization, the itemset-based summarizer must prepare the input document for the subsequent steps. At first, it removes those parts of the text unlikely to be of any value for the final summary. We can specify these parts according to the input text and its logical structure. Since our evaluation corpus consists of a set of biomedical scientific papers, the summarizer removes the figures and tables of the input document as they are not included in the text summarization steps. It will add the figures and tables to the summary in the last step if required. Every other section of the paper that seems to be unnecessary can be removed from the text. Note that the preprocessing step is not merely limited to scientific articles. This action is customizable regarding the logical structure of the input text and the requirements of users.

After removing the unnecessary parts, the summarizer maps the plain text to biomedical concepts contained in the UMLS Metathesaurus to ease the creation of the concept-based summarization model. For this purpose, we use the MetaMap program [43] developed by the National Library of Medicine. Each extracted concept belongs to a subject category, called semantic type. The semantic types divide the concepts into broader categories such as Biologic Function, Cell, Disease or Syndrome, Finding, Sign or Symptom, Mental Process, or Genetic Function. The UMLS defines over 130 semantic types and a set of semantic relations among them in its Semantic Network.

MetaMap uses NLP and computational linguistic methods to match phrases and concepts [43]. It assigns a score to each mapping between a noun phrase and its paired concepts, returning the highest scoring concept along with its semantic type. We use MetaMap with its Word Sense Disambiguation (WSD) feature (i.e. –y flag). Using this option, when multiple mappings with similar scores are returned for a concept, MetaMap tries to resolve the ambiguity and utilizes the Journal Descriptor Indexing (JDI) algorithm [44]. This algorithm attempts to select a single mapping according to the context in which the phrase appears. However, in some cases, it fails to select a single mapping, and MetaMap returns multiple candidate concepts with equal scores. In this case, our method considers all the mappings returned for a phrase. Plaza et al. [45] have shown that the *all mapping* strategy can perform well in WSD for concept-based biomedical summarization as it obtains results comparable to those reported by the best-evaluated WSD methods. For the itemset-based summarizer, we use the 2016 version of the MetaMap program for the mapping step and the 2015AB release of UMLS as the knowledge base. We use MetaMap by the following list of options: -V USAbase -L 15 -Z 1516 -E -Ay --XMLf –E.

Fig. 2 shows a sample sentence and its extracted concepts from the UMLS Metathesaurus. Plaza et al. [5] have identified nine excessively broad semantic types whose concepts can be discarded because they are considered as very generic. These semantic types include *functional concept*, *quantitative concept*, *qualitative concept*, *spatial concept*, *temporal concept*, *language*, *idea or concept*, *intellectual product*, and *mental process*. After mapping the text to the concepts, the itemset-based summarizer discards concepts belonging to these nine semantic types. Therefore, the following concepts are discarded in the sentence represented in Fig. 2: *Widening*, *analysis aspect*, *Further*, *Relationships*, and *Etiology aspects*.

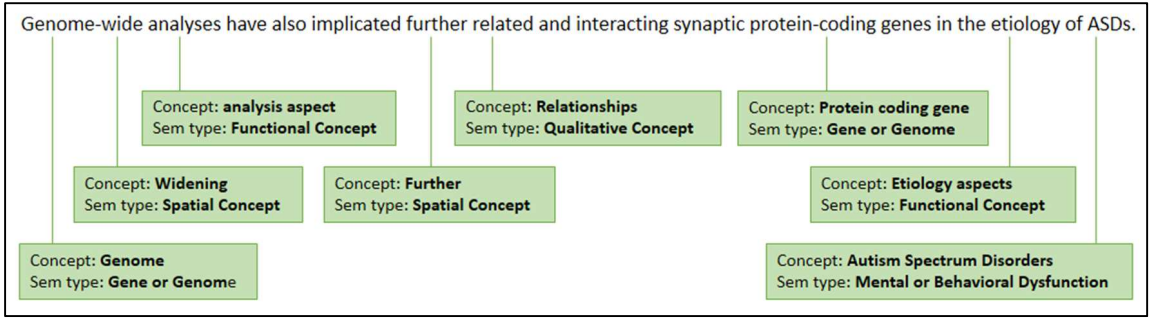

**Fig. 2.** A sample sentence and its extracted concepts from the UMLS Metathesaurus.

### 3.1.2. Data preparation for itemset mining

After the mapping phase and concept extraction, we should represent the input document in an appropriate format for the itemset mining step. The most common data format to represent the input data in itemset and association rule mining is transactional data format [13]. A transactional dataset consists of a set of transactions where each transaction contains a set of items. We should convert the input dataset to the transactional format using a proper approach, according to the context in which itemset mining is used. At this step, we have a set of sentences $\{s_1, …, s_k\}$ where each sentence $s_i$ contains a number of concepts extracted in the previous step. The summarizer converts the set of sentences and their concepts to the transactional data format by considering each sentence $s_k$ as a transaction $tr_k$. It also considers the distinct concepts appearing in $s_k$ as the items of $tr_k$. A more formal definition is given below.

**Definition 1** (*Transactional representation of a document*). Let $D$ be an input document. The transactional representation $T$ of $D$ includes a set of transactions, where each transaction $tr_k$ corresponds to a sentence $s_k$, such that $s_k \in D$, and $tr_k$ consists of distinct items $i_j$, where $i_j$ is $j$th concept in $s_k$.

Based on the above definition, we convert the input document $D$ to transactional representation $T$. For example, Table 1 represents three sentences of a sample document[2]. Table 2 gives the transactional representation of these sentences. We assign a transaction number to each transaction that refers to the corresponding sentence. The distinct concepts extracted from the sentences make the items. In the itemset mining step, the summarizer will use the transactional representation of the input document to generate a concept-based model based on latent correlations among the concepts (items).

---

[2] Available at http://genomemedicine.biomedcentral.com/articles/10.1186/gm102

**Table 1**
Three sentences of the sample document before preparation for itemset mining.

| Sentence number | Content |
| --- | --- |
| 2 | "Genetic epidemiological studies of autism, bipolar disorder and schizophrenia show that the risk of developing one of these specific psychiatric illnesses is proportional to the amount of genetic material shared with an affected individual." [46] |
| 9 | "The distinction between schizophrenia and bipolar disorder has been justified for many years by reference to family studies showing that these disorders seem to 'breed true'." [46] |
| 23 | "Genome-wide analyses have also implicated further related and interacting synaptic protein-coding genes in the etiology of ASDs." [46] |

**Table 2**
Three transactions corresponding to the three sentences represented in Table 1, after preprocessing, mapping and preparation steps. The non-essential concepts are discarded, and the remaining concepts make the items.

| Transaction number | Items |
| --- | --- |
| 2 | Study of Epidemiology, Autistic Disorder, Bipolar Disorder, Schizophrenia, Mental disorders, Genetic Materials, Persons |
| 9 | Schizophrenia, Bipolar Disorder, Reference Object, family investigation, Family, Disease, Breeding |
| 23 | Genome, Protein coding gene, Autism Spectrum Disorders |

### 3.1.3. Extracting main subtopics

At this stage, the itemset-based summarizer discovers the recurrent combinations of concepts from the transactional representation of the input document. The output of this step is a set of frequent *k*-itemsets that demonstrate the themes of the text. A *k*-itemset is an itemset with length of *k* that contains a set of *k* distinct items. If all items of an itemset appear in a given transaction $tr_k$, i.e. the transaction corresponding to sentence $s_k$, it can be said that the itemset covers $tr_k$. In itemset mining, there is a property that measures the frequency of each itemset in proportion to the total number of transactions in the dataset. This property is called *support* and for each itemset is defined as follows:

**Definition 2** (*Itemset support*). Given the transactional representation *T* corresponding to document *D* and an itemset *I*, the value of support property for itemset *I* in *T* is determined as the number of transactions in *T* that are covered by *I*, divided by the total number of transactions in *T*.

For example, in the sample document with 85 sentences, {Proteins} is a 1-itemset whose corresponding concept appears in seven sentences. Therefore, the {Proteins} covers seven transactions in *T*, and its support value is $\frac{7}{85}$. The {Deletion Mutation, NRXN1 gene} is a 2-itemset and the {Autistic Disorder, Bipolar Disorder, Schizophrenia} is a 3-itemset. These two sets cover six and nine transactions in *T*, respectively. Hence, their support values are $\frac{6}{85}$ and $\frac{9}{85}$. The itemset-based summarizer considers the support values of itemsets as a metric to measure their significance.

In the itemset mining step, the algorithm produces a large number of itemsets. However, just the *frequent itemsets* are useful for the subsequent steps. In the following, we present a formal definition of a frequent itemset.

**Definition 3** (*Frequent itemset*). Itemset *I* is said to be frequent in the transactional representation *T* of document *D*, if its support value in *T* is greater than or equal to a given minimum support threshold *min_sup*.

For example, let the *min_sup* value for the sample document be equal to $\frac{7}{85}$. Based on this value, the early mentioned itemsets {Proteins} and {Autistic Disorder, Bipolar Disorder, Schizophrenia} are frequent itemsets, and the itemset {Deletion Mutation, NRXN1 gene} is not frequent.

Given a transactional representation *T* of document *D* and a minimum support threshold *min_sup*, the summarizer executes the frequent itemset mining process to discover all the frequent itemsets in *T*. To perform the frequent itemset mining task, we employ the Apriori [13, 14], a basic algorithm in data mining. The Apriori algorithm is basically used for association rule mining. In our proposed method, we utilize it just to extract the frequent itemsets. We do not use its ability to generate association rules. The itemset-based summarizer considers the extracted frequent itemsets as the main themes of the input text and utilizes them to select the most related and informative sentences. The inputs of the algorithm are the transactional representation *T* of document *D*, and the minimum support threshold *min_sup*. The output is a set of discovered frequent itemsets *FI* where each itemset demonstrates a subtopic of text and contains a set of correlated and recurrent concepts.

**Table 3**
The frequent itemsets extracted from the transactional representation of the sample document related to 'the genetic overlap between autism, schizophrenia, and bipolar disorders'. The itemsets are sorted based on their support values.

| Itemset | Support | Itemset | Support |
| --- | --- | --- | --- |
| {Schizophrenia} | 0.352 | {Study} | 0.094 |
| {Bipolar Disorder} | 0.235 | {Tryptophanase} | 0.094 |
| {Bipolar Disorder, Schizophrenia} | 0.223 | {Disease} | 0.094 |
| {Autism Spectrum Disorders} | 0.223 | {Binding (Molecular Function)} | 0.082 |
| {Autistic Disorder} | 0.200 | {Scientific Study} | 0.082 |
| {neuroligin} | 0.176 | {Reporting} | 0.082 |
| {Deletion Mutation} | 0.164 | {Proteins} | 0.082 |
| {Autistic Disorder, Schizophrenia} | 0.152 | {neuroligin, Binding (Molecular Function)} | 0.070 |
| {Genes} | 0.129 | {Deletion Mutation, NRXN1 gene} | 0.070 |
| {NRXN1 gene} | 0.129 | {Schizophrenia, Autism Spectrum Disorders} | 0.070 |
| {Autistic Disorder, Bipolar Disorder, Schizophrenia} | 0.105 | {Mental Retardation} | 0.070 |
| {Autistic Disorder, Bipolar Disorder} | 0.105 | {Procedure findings} | 0.070 |
| {Copy Number Polymorphism} | 0.105 | {Staphylococcal Protein A} | 0.070 |
| {Genome} | 0.105 | {Encode (action)} | 0.070 |
| {Persons} | 0.105 | {Genome-Wide Association Study} | 0.070 |
| {Alleles} | 0.094 | {Diagnosis} | 0.070 |

Table 3 represents the frequent itemsets extracted from the sample document. The sample document is about 'the Genetic overlap between autism, schizophrenia, and bipolar disorders'. In this example, the itemset mining algorithm extracts a total number of 32 frequent itemsets, 25 of which are 1-itemsets, six of which are 2-itemsets, and one is 3-itemset. In this example, the value of minimum support threshold *min_sup* is 0.07, i.e. an itemset is frequent if it covers at least 7% of the transactions. In other words, a set of correlated concepts is a central subtopic of the text if the concepts appear together in at least 7% of the sentences. It is shown in Table 4 that the average number of sentences of a document in our evaluation corpus is 167. According to this average number, it can be said, for a minimum support threshold of 0.07, in average an itemset must appear at least in 12 sentences to be discovered as a frequent itemset. This *min_sup* value of 0.07 used in the above example is not the optimum value for this threshold. The optimum value of *min_sup* threshold is specified by a set of preliminary experiments in Section 4.1.1 and is discussed in Section 5.1.

### 3.1.4. Sentence scoring and summary creation

After performing the itemset mining step that discovers the main subtopics, we should rank the sentences of the input text based on how much they are informative and cover the main subtopics. Hence, we require a scoring strategy to quantify the informativeness of the sentences.

The support value of each itemset indicates that how much the itemset is significant. It means that in the comparison between two itemsets extracted from a given document, an itemset is assumed to be more valuable and informative if it has a higher support value. Therefore, we hypothesize that the support value of the itemsets that cover a sentence is an appropriate measure to quantify the importance of the sentence. The itemset-based summarizer assigns a score to each sentence by adding the support value of the itemsets that cover the sentence. The score of each sentence is calculated as follows:

$$Score(s_i) = \sum_{FI_j \mid FI_j \in FI \wedge FI_j \text{ covers } tr_i} Support(FI_j) \qquad (1)$$

where $s_i$ is $i_{th}$ sentence in document $D$, and $tr_i$ is the transaction corresponding to $s_i$. $FI$ is the set of all frequent itemsets extracted from the transactional representation $T$ of document $D$, and $FI_j$ is $j_{th}$ itemset in $FI$.

After scoring all the sentences of the input document, the summarizer can decide which sentences should be selected to be included in the final summary. Two factors contribute to a higher score: 1) the number of frequent itemsets that cover the sentence, and 2) the support value of the frequent itemsets that cover the sentence. Therefore, the sentences that contain more frequent itemsets and are covered by high supporting frequent itemsets produce higher scores. The summarizer considers the high-scoring sentences to be highly informative and more related to the main topics of the input text. After the sentence-scoring task, the summarizer sorts the sentences based on their scores. It selects the top *N* sentences to build the final summary, where *N* is the number of sentences that must be chosen for the summary and is determined by the compression rate. If two sentences have the same score, the method assigns a higher priority to the sentence having shorter length. The summarizer arranges the selected sentences according to their appearance order in the input text. Finally, if the summary refers to a figure or a table of the primary document removed in the preprocessing step, the method adds the figure or the table to the summary. Fig. 3 shows the summary of the sample document produced by the itemset-based summarizer. In this example, for brevity reasons, the compression rate is 10%. That is, the size of the summary must be 10% of the input document.

> Such diagnostic categories are therefore likely to be heterogeneous and the boundaries between them somewhat arbitrary. Autism, schizophrenia and bipolar disorder have traditionally been considered as separate disease entities, although they do share some common behavioral characteristics and cognitive deficits.
> Therefore, just as for NRXN1 deletions, it is apparent that these large CNVs confer risk of a range of neurodevelopmental phenotypes, including autism, mental retardation and schizophrenia.
> The advent of the GWAS has allowed most of the common SNP variation in the human genome to be tested for association and the first wave of such studies has been reported for schizophrenia, bipolar disorder and autism.
> Furthermore, there have been recent reports of association for common alleles at several GABA receptor genes in a subtype of bipolar disorder and schizophrenia, which implicate loci also reported as associated with ASDs.
> Such biological roles fit with hypotheses of the etiology of autism and schizophrenia in which a neurodevelopmental insult and adult imbalance in excitatory and inhibitory neurotransmission occur in the absence of overt macro-pathology. SHANK3 is implicated in autism by several lines of evidence and functions as a post-synaptic scaffolding protein that binds indirectly to neuroligins, forming a potentially functional circuit of neurexin-neuroligin-Shank that is dysregulated in ASDs.
> Whole-genome studies of many thousands of affected individuals are uncovering evidence for genetic overlap between autism, schizophrenia and bipolar disorder.
> Studies of CNVs and other rare alleles have found overlap between autism and schizophrenia, whereas those of common SNP variants have shown overlap between schizophrenia and bipolar disorder.
> The findings also support the view that schizophrenia has a stronger neurodevelopmental component than bipolar disorder and suggest that it lies on a gradient of decreasing neurodevelopmental impairment between syndromes such as mental retardation and autism, on one hand, and bipolar disorder on the other.
> We have based this conclusion on the fact that several rare CNVs, including deletions of NRXN1, are associated with mental retardation, autism and schizophrenia, and on the overlap in common risk alleles seen between schizophrenia and bipolar disorder.

**Fig. 3.** The summary of the sample document generated by the itemset-based summarizer (compression rate=10%).

### 3.2. Evaluation method

The performance of text summarization systems can be evaluated using two methods, *Extrinsic* and *Intrinsic* [47]. The extrinsic method evaluates the impact of summarization on the quality of specific tasks that use produced summaries. As some of these specific tasks we can point out those performed by question answering systems or search engines [1]. The intrinsic method evaluates the performance of summarization regarding the measures that assess the quality of summaries. The most common measures used for intrinsic evaluation are informativeness, accuracy, relevancy, comprehensiveness, and readability [1]. Since we intend to assess the competency of the itemset-based summarizer regarding the informative content of produced summaries, we evaluate its performance through intrinsic evaluation.

**Table 4**
The statistics of the evaluation corpus containing 400 biomedical scientific papers and their abstracts. The minimum, maximum, and average number of sentences and words of a document are presented for the full-texts and the abstracts.

|  | **Number of sentences** | | | **Number of words** | | |
| --- | --- | --- | --- | --- | --- | --- |
|  | **Min.** | **Max.** | **Average** | **Min.** | **Max.** | **Average** |
| **Full-texts** | 54 | 398 | 167 | 1003 | 9789 | 3917 |
| **Abstracts** | 7 | 19 | 12 | 109 | 376 | 251 |

### 3.2.1. Evaluation dataset and metrics

To assess the performance of the proposed summarization method, we need to summarize a corpus of biomedical text documents that have model summaries available (to serve as a reference standard). The model summary is used to measure the similarity between its content and the system-generated summary. To the authors' knowledge, such a corpus has not been developed for the evaluation of single-document biomedical text summarization methods. However, a common approach is summarizing a collection of biomedical articles, and using the abstract of each article as the model summary [5]. To evaluate the performance of the itemset-based summarizer, we employ a collection of 400 biomedical articles, randomly selected from BioMed Central's corpus for text mining research[3]. We use the abstracts of the papers as the model summaries. This dataset is large enough for the results of the evaluations to be significant [48]. Table 4 presents the statistics of the evaluation corpus.

To evaluate the performance of the itemset-based summarizer in terms of its ability to improve the informativeness of automatic summaries, we use the ROUGE package [15]. ROUGE estimates the shared content by comparing a system-generated summary with single or multiple model summaries. The estimation is performed by calculating the proportion of shared *n*-grams between the system and model summaries. For the evaluations, we use four ROUGE metrics:

- ROUGE-1 (R-1): It computes the number of shared unigrams between the system and model summaries.
- ROUGE-2 (R-2): It computes the number of shared bigrams between the system and model summaries.
- ROUGE-W-1.2 (R-W-1.2): It compares the consecutive matches between the system and model summaries and computes the union of the longest common subsequences.
- ROUGE-SU4 (R-SU4): It calculates the overlap of skip-bigrams (pairs of words having intervening word gaps) between the system and model summaries and allows a skip distance of four between the bigrams.

---

[3] http://old.biomedcentral.com/about/datamining

The ROUGE metrics compute scores between 0 and 1. A higher score indicates greater content overlap between the system and model summaries. Therefore, a summarizer is assumed to be better if its produced summaries obtain higher ROUGE scores, because they contain more shared content with the model summaries and are considered to be more informative. It is worth mentioning that the ROUGE scores showed high correlation with the human judges of the Document Understanding Conference (DUC) evaluation data [15], and DUC conference adopted it as the official evaluation metric for text summarization.

### 3.2.2. The value of minimum support threshold

We conduct two sets of experiments to evaluate the performance of our itemset-based method for biomedical text summarization. To tune the parameter of the system, the first part of experiments is described in this subsection. Afterward, the second part of experiments will be described in the next subsection that evaluates the itemset-based summarizer against other summarization methods.

We design and perform a set of preliminary experiments in order to determine the best value for the *min_sup* threshold involved in discovering frequent itemsets (Section 3.1.3). The Apriori algorithm finds the frequent itemsets whose support value is greater than or equal to the *min_sup* threshold. Then, the itemset-based summarizer considers the extracted frequent itemsets to be the main themes of the document. It assigns a score to each sentence based on the frequent itemsets that cover the sentence. Therefore, the higher the value of *min_sup* threshold, the fewer the frequent itemsets to be discovered and the fewer the number of itemsets participating in the sentence scoring step. In contrast, the lower value of *min_sup* threshold, the more the frequent itemsets to be extracted and the more the itemsets getting involved in the sentence scoring phase. We assess the impact of lower and higher values of the *min_sup* threshold on the performance of the proposed itemset-based summarizer. We choose the threshold value reporting the highest ROUGE scores as the best value for the *min_sup*. We use the optimum value in the subsequent experiments to evaluate the itemset-based summarizer against other summarization methods.

To specify the best value for the *min_sup* threshold as well as for parameterization of the comparison methods, we use a separate evaluation corpus consisting of 100 biomedical articles, randomly selected from BioMed Central's corpus for text mining research. We use the abstracts of the articles as the model summaries.

### 3.2.3. Comparing with other summarizers

We compare our proposed summarization method against other research-oriented, publicly available, commercial, and baseline summarizers to assess its appropriateness for biomedical literature summarization. The other summarizers used in the evaluations include *GraphSum* [39], *TexLexAn* [49], *SweSum* [21], *SUMMA* [9], *Microsoft AutoSummarize*, *the term-based version of the itemset-based summarizer*, *Lead baseline*, and *Random baseline*. The rationale for the selection of these comparison methods is the utilized term-based methods and generic measures. Since the challenges we address in this study are not related to previous biomedical summarizers, we do not use any biomedical summarization methods (mentioned in Section 2) in our evaluations and comparisons. For example, the semantic graph-based approach proposed by Plaza et al. [5] is one of the few methods concerning biomedical literature summarization. It addresses the challenges related to the previous graph-based methods by considering concepts and their relations as the nodes and edges of the graph. If we include this method in our evaluations, no matter which system would obtain better scores, the difference between the semantic graph-based system and our summarizer could not be discussed to address the challenges introduced in Section 1. In the following, the comparison methods are described.

GraphSum [39] is a domain-independent summarizer that uses a graph-based approach in combination with association rule mining. It creates a correlation graph where the nodes represent sets of recurrent terms extracted regarding frequent itemsets, and the edges indicate positive and negative correlations among multiple terms in pairs of nodes. It generates the summary using a greedy strategy, selecting the most representative subset of sentences that best cover the correlation graph.

TexLexAn [49] is a widely used and open source automatic summarizer that uses keywords extracted from the input text and the cue expressions to assess the relevancy of sentences and extract the most relevant ones for the final summary. SweSum [21] is a research-oriented summarizer whose online version is available for public usage and is considered to be state-of-the-art for English, Danish, Norwegian and Swedish text summarization. For sentence scoring, it uses a set of features, such as presence in the first line of the text, presence of numerical values in the sentence, and the presence of keywords in the sentence. SweSum provides an option to choose between 'Academic' and 'Newspaper' as the type of text. For this option, we use the 'Academic' text type. SUMMA [9] is another research-oriented summarizer for both single and multi-document summarization. It selects the sentences to be included in the summary based on several statistical and similarity-based features. Each feature has a weight that specifies its importance. For the evaluations, we use these features for SUMMA, i.e. the frequency of terms contained in the sentence, the position of the sentence in the document, the similarity of the sentence to the first sentence of the document, and the overlap between the sentence and the title of the document. AutoSummarize is a commercial application and a feature of the Microsoft Word software [50]. It performs the sentence scoring and summarization tasks based on a word frequency algorithm. Higher scores are assigned to the sentences that contain more frequent words. We use these four summarizers to assess the competency of statistical and similarity feature-based, keyword-based, and word frequency summarization methods for biomedical literature summarization.

The term-based version of the itemset-based summarizer uses terms instead of concepts to build an itemset-based model. In the term-based version of our method, the stage of mapping text to biomedical concepts is no longer needed, but two additional steps are added for extracting appropriate terms that can be used in the itemset mining stage. The first step is stop-word elimination that removes highly frequent words that have little lexical content. To this aim, we employ the Natural Language Toolkit (NLTK) stop-word corpus [51]. The second step is stemming, reducing all the words contained in the sentences to their corresponding stem. For this purpose, we adopt the porter stemming algorithm [52]. After applying these two steps, the remaining terms form the items. The summarizer performs the remaining steps similar to the original concept-based version. The goal of using the term-based version for the evaluations is to investigate the impact of concept-level analysis of text on the accuracy of the itemset-based text modeling against term-level analysis.

We also use two baseline methods in the experiments. The Lead baseline method returns the first $N$ sentences of the input text as the summary, and the Random baseline randomly selects $N$ sentences and generates the summary.

We also perform a set of preliminary experiments to parameterize the comparison methods containing tunable parameters on the same development set employed for the parameterization of the itemset-based summarizer. The aim of parameterization of all the methods on the same development set is to validate the results of the final evaluations where the performance of the methods is investigated for biomedical literature summarization. The results of the parameter tuning will be presented in Section 4.1.2.

In our evaluations, we set the compression rate to 30% for all the summarizers. That is, the size of the generated summary is equal to 30% of the input document. This choice is based on a widely accepted standard of a compression rate between 15% and 35% [53]. We assess the statistical significance of the results using a Wilcoxon signed-rank test with a 95% confidence interval.

## 4. Results

In this section, first, we present the results of the preliminary experiments and the parameterization of the itemset-based summarizer and the comparison methods. Then, the results of evaluations comparing the itemset-based summarizer with the other summarization methods will be presented.

### 4.1. Parameterization results

We perform the first set of parameter tuning experiments on the development corpus mentioned in Section 3.2.2 to assess the impact of the *min_sup* threshold on the performance of the itemset-based summarizer. We also devote the second set of preliminary experiments to the parameterization of the competitor methods. In the following, we separately present the results of these two sets of experiments.

#### 4.1.1. Itemset-based summarizer

The Apriori algorithm decides whether the extracted itemsets are frequent according to the *min_sup* threshold. As mentioned in Section 3.2.2, the number of discovered frequent itemsets has an inverse relationship with the value of the support threshold. In fact, the higher the value of the minimum support threshold, the fewer the number of frequent itemsets to be discovered and vice versa. Fig. 4 presents the results of the experiments. For brevity reasons, only R-2 and R-SU4 scores are reported. The best value for the *min_sup* threshold is 0.08 (i.e. 8%) that provides the best scores for the ROUGE metrics (R-2: **0.3359**, and R-SU4: **0.3884**). For the final evaluations, according to the average number of sentences of a document in the evaluation corpus (Table 4), it can be said that an itemset must approximately appear at least in 14 sentences to be discovered as a frequent itemset for a minimum support threshold of 0.08.

It can be observed from Fig. 4 that, according to the ROUGE scores, the value of *min_sup* threshold has a significant impact on the quality of the generated summaries. Table 5 reports the average number of all discovered frequent itemsets and the average number of *k*-itemsets, where *k* is equal to 1, 2, 3, and 4. The average numbers are given for each tested support threshold.

We perform similar experiments to specify the best value of the *min-sup* threshold for the term-based version of the itemset-based summarizer. The threshold value of 0.1 reports the highest ROUGE scores (R-2: **0.3286**, and R-SU4: **0.3721**), and we use this value for the subsequent experiments.

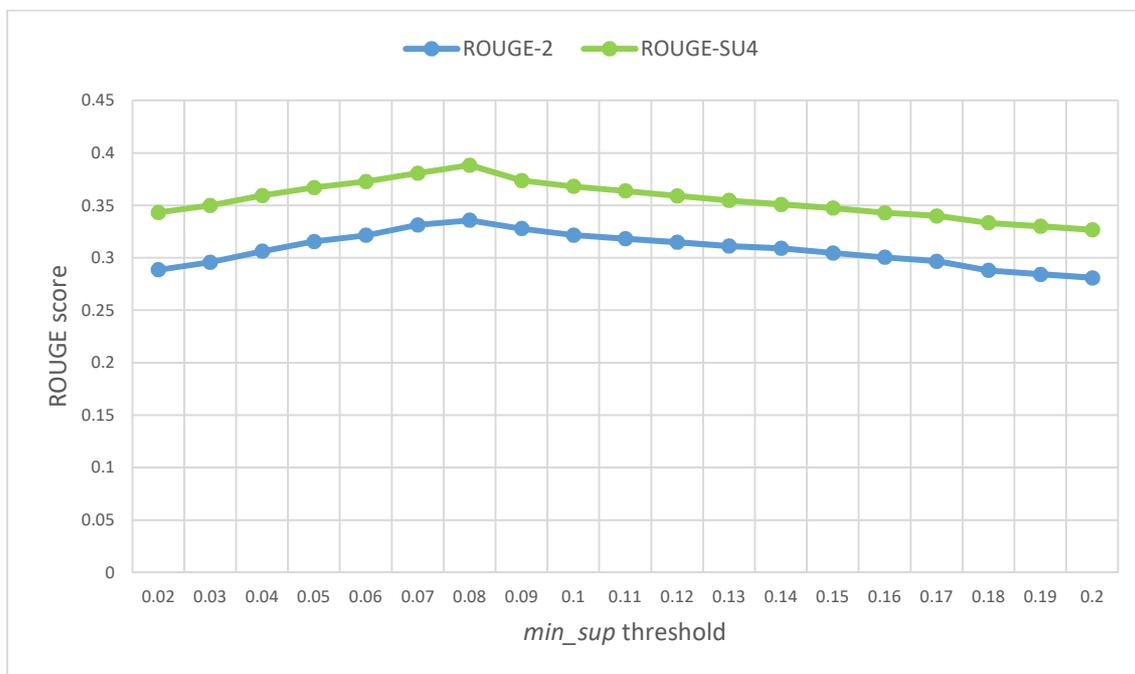

**Fig. 4.** ROUGE-2 and ROUGE-SU4 scores for different values of the minimum support threshold *min_sup*.

**Table 5**
The average number of all extracted frequent itemsets and the average number of *k*-itemsets, for the preliminary experiments evaluation corpus. The average numbers are given for each tested minimum support threshold in the preliminary experiments. The numbers are rounded off to one decimal digit.

| *min_sup* threshold | Average number of all frequent itemsets | Average number of frequent 1-itemsets | Average number of frequent 2-itemsets | Average number of frequent 3-itemsets | Average number of frequent 4-itemsets |
|---|---|---|---|---|---|
| **0.02** | 1891.0 | 96.2 | 239.0 | 355.4 | 416.6 |
| **0.03** | 974.9 | 62.6 | 129.0 | 179.2 | 200.8 |
| **0.04** | 557.4 | 46.4 | 76.9 | 100.0 | 110.9 |
| **0.05** | 322.0 | 36.2 | 52.3 | 63.0 | 62.9 |
| **0.06** | 166.1 | 27.5 | 33.2 | 34.1 | 30.2 |
| **0.07** | 133.0 | 22.8 | 25.9 | 25.4 | 23.0 |
| **0.08** | 74.3 | 18.9 | 19.0 | 16.7 | 11.9 |
| **0.09** | 60.1 | 15.5 | 15.0 | 13.1 | 9.6 |
| **0.10** | 51.5 | 12.8 | 12.5 | 11.1 | 8.5 |
| **0.11** | 35.8 | 11.3 | 10.0 | 7.6 | 4.6 |
| **0.12** | 23.9 | 9.6 | 7.8 | 4.5 | 1.7 |
| **0.13** | 19.2 | 7.7 | 6.4 | 3.7 | 1.2 |
| **0.14** | 16.5 | 6.9 | 5.3 | 3.1 | 1.1 |
| **0.15** | 15.2 | 6.3 | 4.9 | 2.9 | 1.0 |
| **0.16** | 12.5 | 5.4 | 4.0 | 2.3 | 0.8 |
| **0.17** | 11.8 | 5.0 | 3.8 | 2.2 | 0.8 |
| **0.18** | 9.8 | 4.3 | 3.1 | 1.8 | 0.5 |
| **0.19** | 6.8 | 3.6 | 2.1 | 0.8 | 0.2 |
| **0.20** | 6.2 | 3.2 | 2.0 | 0.8 | 0.2 |

### 4.1.2. Comparison methods

In this section, we present the results of parameterization of those competitor summarizers containing tunable parameters. GraphSum method contains three parameters that can affect the quality of produced summaries. These parameters include 1) the minimum support threshold to extract frequent itemsets, 2) the minimum lift threshold to identify positive correlations between terms, and 3) the maximum lift threshold to identify negative correlations between terms. We experiment GraphSum by varying the minimum support threshold in the range [0.02, 0.1], the minimum lift threshold in the range [5, 35], and the maximum lift threshold in the range [0.1, 0.9]. The minimum support threshold of 0.05, the minimum lift threshold of 19, and the maximum lift threshold of 0.4 are the optimum configuration. Using these parameter values, GraphSum reports its best scores (R-2: **0.3381**, and R-SU4: **0.3814**).

SUMMA method scores the sentences of an input document using a set of features and their weights. We tune the parameters of SUMMA varying the weight of each feature and assessing the obtained ROUGE scores under each configuration. The best scores (R-2: **0.3126**, and R-SU4: **0.3511**) are reported when the weights of the term frequency feature, the position feature, the similarity to the first sentence feature, and the overlap with the title feature are 0.35, 0.3, 0.2, and 0.15, respectively. We perform another set of parameterization experiments to find the optimal weights used by SweSum to score the sentences of an input document. The best scores (R-2: **0.3002**, and R-SU4: **0.3427**) are reported when the weights of the first line feature, the numerical values feature, and the keywords feature are 0.2, 0.15, and 0.65, respectively.

For the final evaluations, we run GraphSum, SUMMA, and SweSum using the optimal parameter values specified in the above experiments.

### 4.2. Evaluation results

To evaluate the performance of our proposed itemset-based biomedical summarizer, we compare the ROUGE scores assigned to our method with the scores assigned to the other summarizers, including *GraphSum*, *TexLexAn*, *SweSum*, *SUMMA*, *Microsoft AutoSummarize*, *Itemset-based summarizer (term-based version)*, *Lead baseline*, and *Random baseline*. The ROUGE toolkit assigns four different scores to each summarizer while comparing the system-generated summaries with the model summaries. Table 6 shows the ROUGE scores obtained by all the summarizers. As can be observed, our proposed itemset-based biomedical summarizer reports higher ROUGE scores than the other summarizers and the baseline methods. Compared with GraphSum, the improvement in the scores obtained by the itemset-based summarizer is significant for R-1 score according to Wilcoxon signed-rank test ($p < 0.05$). Compared with the other summarizers, the itemset-based summarizer significantly improves all the reported ROUGE metrics ($p < 0.05$).

The term-based version of the itemset-based summarizer substantially performs better than AutoSummarize, Lead baseline, and Random baseline for all the reported ROUGE scores ($p < 0.05$). Its improvement in the scores, compared with SUMMA, is significant for R-1 and R-SU4 ($p < 0.05$). It also greatly improves R-1, R-2, and R-SU4 scores with respect to SweSum and TexLexAn, although, its improvement is not significant for R-W-1.2 score ($p > 0.05$). Its obtained ROUGE scores are lower than those of GraphSum, and the difference is significant for R-1 and R-2 scores ($p < 0.05$).

**Table 6**
ROUGE scores obtained by the itemset-based summarizer and the comparison methods. The best score for each metric is shown in bold type. The summarizers are sorted based on their ROUGE-2 scores.

|  | ROUGE-1 | ROUGE-2 | ROUGE-W-1.2 | ROUGE-SU4 |
|---|---|---|---|---|
| *Itemset-based summarizer* | **0.7583** | **0.3381** | **0.0934** | **0.3889** |
| *GraphSum* | 0.7460 | 0.3359 | 0.0913 | 0.3795 |
| *Itemset-based summarizer (term-based)* | 0.7297 | 0.3219 | 0.0853 | 0.3694 |
| *SUMMA* | 0.7164 | 0.3178 | 0.0829 | 0.3548 |
| *SweSum* | 0.6983 | 0.3014 | 0.0796 | 0.3462 |
| *TexLexAn* | 0.6891 | 0.2956 | 0.0784 | 0.3420 |
| *Lead baseline* | 0.6322 | 0.2529 | 0.0721 | 0.3196 |
| *AutoSummarize* | 0.6281 | 0.2436 | 0.0692 | 0.3125 |
| *Random baseline* | 0.5602 | 0.2128 | 0.0669 | 0.2908 |

## 5. Discussion

In this section, we discuss the results of the parameterization experiments and the final evaluations presented in Section 4.

### 5.1. Parameterization

As reported in Table 5 and shown in Fig. 4, when we select a relatively small *min_sup* value, e.g. 0.02 or 0.03, the itemset mining algorithm returns a large number of frequent itemsets. These itemsets participate in the sentence scoring step whereas only a small number of them contain useful information to assess the informative content of the sentences. As a result, the summarizer performs the sentence scoring step using numerous redundant itemsets. This negatively affects the quality of the produced summaries. For example, when the value of *min_sup* threshold is 0.02, the method extracts a total number of 477 frequent itemsets from the sample document mentioned in Section 3. The number of *k*-itemsets for $k = 1, 2, 3, 4, 5$, and 6 is equal to 85, 204, 125, 48, 13, and 2, respectively. The results show that the produced summary tends to cover a range of concepts including *Reproductive History*, *Study of Epidemiology*, *Population Group*, *Sharing*, *Application procedure*, *Comparison*, *Social Role*, etc when such a large number of itemsets participate in sentence scoring. It seems that such concepts are weakly relevant to the main subtopics as they appear less frequently in the text. In this particular example, the itemset-based summarizer reaches R-2: **0.2903** and R-SU4: **0.3459** scores.

As another extreme threshold, when we choose a relatively high value, e.g. 0.19 or 0.2, for the value of *min_sup*, the number of discovered frequent itemsets is remarkably reduced. In this case, the summarizer quantifies the informativeness of the sentences according to a low number of itemsets whereas this number of itemsets is not enough to decide to what extent the sentences are informative. In fact, the summarizer assigns scores to the sentences based on a limited number of main subtopics while there are other important subtopics disregarded by an extreme value of the support threshold. Consequently, the summarizer cannot accurately assign scores to the sentences based on their actual informativeness. Hence, the quality of the produced summaries is lowered. For example, when the value of *min_sup* threshold is 0.19, the method extracts a total number of five frequent itemsets from the sample document. There are four frequent 1-itemsets and one frequent 2-itemset. The results show that the summary extremely tends to cover the most frequent concepts, like *Autism Spectrum Disorders*, *Bipolar Disorder*, and *Schizophrenia* when this limited number of itemsets are used for sentence scoring. In this particular example, the itemset-based summarizer receives R-2: **0.2871** and R-SU4: **0.3318** scores.

On the other hand, the itemset mining algorithm extracts a total number of 23 frequent itemsets from the sample document when the value of *min_sup* threshold is 0.08. There are 19 frequent 1-itemsets, three frequent 2-itemsets, and one frequent 3-itemset. The results show that the summary covers both the most important concepts demonstrating the primary subtopics and other relevant concepts including *Genome*, *Proteins*, *Deletion Mutation*, *Alleles*, *neuroligin*, *Binding (Molecular Function)*, and *NRXN1 gene* using these extracted itemsets for sentence scoring. In this case, compared to the summary generated by a small threshold of 0.02, the concepts that seem to be weakly relevant to the main subtopics appear less frequently in the summary. Moreover, compared to the summary generated by a high threshold of 0.19, the extreme tendency of the summary to cover the main subtopics is lowered and other relevant subtopics appear more frequently. In this particular example, the itemset-based summarizer obtains R-2: **0.3407** and R-SU4: **0.3915** scores.

For GraphSum, which also employs itemset mining, it also has been shown that an optimum support threshold plays an important role in balancing model specialization and generality [39].

### 5.2. Comparing with other summarizers

Investigating the summaries produced by different methods, we can point out the following observations and discuss the evaluation results presented in Table 6.

The results show that the Random baseline performs worse than the other summarizers. As expected, sentences randomly selected from the documents do not lead to producing useful summaries. The Lead baseline performs better than the Random baseline because a portion of valuable information of a scientific article is presented when beginning with the introduction.

However, the summarization performance of the Lead baseline is not satisfying, and it is required all the relevant information throughout the document be contained in the final summary. Moreover, other types of biomedical text documents may not have the same structure as the scientific articles, and the Lead baseline may give worse summarization performance when other types of text documents are used.

The word frequency algorithm of AutoSummarize produces the summaries based on the highly-frequent words contained in the input texts. In comparison with the top ranked summarizers, the summaries generated by this method show a narrower topic coverage. Moreover, the presence of irrelevant and low-informative sentences lowers the quality of the summaries. With the development of more intelligent methods utilizing a variety of features of text [18], it can be argued that the term frequency feature cannot solely be used to produce high-informative summaries, particularly in biomedical literature summarization.

In the summaries produced by SweSum, the sentences containing numerical values appear more frequently. These sentences are mostly extracted from the results section of the documents. Many important sentences concerning the main subtopics of the documents are missing from the summaries. So this summarization method may be more useful when the user prefers to know about the results reported in a biomedical article. Since the model summaries used for the evaluations (the abstract of the articles) address different parts of the articles (background, methods, results, etc.), SweSum reports lower scores compared to the other summarizers which do not differentiate between numerical and other content. However, when the weights of SweSum's features are tuned, the performance of SweSum increases and its tendency to select the sentences containing numerical values decreases. It has been shown that, in biomedical literature summarization, produced summaries convey more informative content when important sentences are selected from different sections of an input document [8].

TexLexAn uses the keywords extracted from the input text to score the sentences. Looking at the summaries generated by this method, we can find out that the extracted keywords are the high-frequency terms in each document. There are some similarities between the summaries of TexLexAn, SweSum, SUMMA, and AutoSummarize. The reason for this similarity can be the dependency of these methods on the high-frequent terms as the keywords of the text. Assessing the documents and summaries, we observe that some keywords extracted by these methods are essentially meaningless or unrelated to the topics of the documents. For example, the following terms are among the keywords extracted from a number of documents investigated by the four summarizers: *those*, *which*, *other*, *their*, *after*, *number*, *using*, *there*, *during*, *could*, *significant*.

The above summarizers utilize such generic terms to measure the relatedness and informativeness of the sentences. Despite their high frequency, the itemset-based summarizer discards such terms from the produced model, since their semantic types are excessively broad (as explained in Section 3.1.1) or they are mapped to no concept in the preprocessing step. We can regard this improvement in selection as a key reason for the better ROUGE scores reported by the itemset-based summarizer. It seems that the investigation of more complicated keyword extraction methods employed in other fields, like text classification [54], could improve the performance of the available summarizers.

SUMMA employs a word frequency method as well as similarity-based and positional features. It obtains better ROUGE scores compared to TexLexAn, SweSum, and AutoSummarize. This better summarization quality could be due to utilizing additional features. The summaries produced by SUMMA show that the similarity of sentences with the title of the document can be a useful feature in selecting more related sentences. This feature has also been investigated for the selection of informative sentences and segments in biomedical text summarization [5, 55]. However, some issues may reduce the effectiveness of this feature. First, all type of input texts may not have a title. Second, a sentence may convey similar information to the title of text, but contains different terms. The produced summaries suggest that this feature could lead to some degree of redundancy because the relatedness of the sentences should be measured by a broader range of topics, not only by the terms appearing in the title.

Another feature implemented by SUMMA is the position of the sentence in the document. Looking at the output summaries, we observe that the first sentences of the paragraphs of a document are more likely to be selected for the summary, although, they may convey no informative content. The summaries show that this feature may not work efficiently for this type of documents where there are numerous paragraphs and the beginning sentences may not necessarily be related to the main topics. However, this sentence selection strategy has shown its usefulness for summarization of newswire articles [19]. Another feature used by SUMMA is the similarity of the sentence to the first sentence of document. The produced summaries show that the sentences having more than five terms in common with the first sentence of the respective document are more likely to be included in the final summary. In the documents with an appropriate first sentence, this feature helps to extract more informative sentences. On the other hand, for the documents with an irrelevant first sentence, this feature leads to some unrelated sentences to be included in the summaries. Regarding the evaluation results, it seems that the methods and features used by SUMMA, TexLexAn, SweSum, and AutoSummarize may not be utilized as adequate measures to quantify the informative content of sentences in this type of summarization.

GraphSum reports the highest ROUGE scores among the term-based summarizers. In comparison with the summaries produced by SUMMA, TexLexAn, SweSum, and AutoSummarize, the output summaries of GraphSum cover a broader range of topics discussed in the input documents. This better coverage of topics could be due to its sentence selection strategy where it tries to select those sentences that best cover the correlation graph of the input text. However, for the relatively large documents (usually more than 150 sentences), GraphSum reports lower ROUGE scores than for the smaller documents. It seems that the accuracy of its hybrid graph-itemset model or its sentence selection method could negatively be affected by complex term distributions and abundant number of positive and negative correlations among terms in large documents [39].

The term-based version of the itemset-based summarizer reports lower scores than GraphSum. Nevertheless, it performs better than the other term-based competitors. It covers a relatively broader range of topics in its produced summaries. However, low-informative and unrelated sentences still appear in the output summaries of the two methods. Based on these observations, it is essential to select the sentences according to their approximated meaning and the subtopics that each sentence covers. The itemset-based summarizer analyzes documents using a concept-based approach to cope with the issues raised by other methods.

The itemset-based summarizer reports the best ROUGE scores. Following the trend of knowledge rich methods in biomedical text summarization [1], the itemset-based summarizer benefits from the combination of domain knowledge and itemset mining based summarization. It can improve the performance of biomedical text summarization by dealing with concepts rather than terms. Through the use of itemset mining, the method extracts correlated concepts in the form of frequent itemsets and considers them as important subtopics. It uses the main subtopics to measure the informative content of the sentences. When the summarizer maps the input text to biomedical concepts, a single concept may be representative of multiple words or phrases sharing a common meaning. This operation can increase the accuracy of the produced model. Moreover, in term-based modeling, the summarizers deal with a sequence of terms indicating a unique concept as separate unit items. On the other hand, using concept-based modeling, our summarizer can deal with a sequence of terms indicating a concept as an unit item. For example, the itemset-based summarizer considers the phrase "*genome-wide association study*" as a unique concept while GraphSum deals with this phrase as three or four separate terms. In this way, the number of items increases and the accuracy of the model generated by GraphSum is reduced.

In our preliminary experiments, we evaluated the comparison methods using the parameter values suggested by the respective authors or their default configurations. For brevity reasons, we present only the ROUGE scores obtained by the optimum parameter values in Section 4.1.2. Comparing the scores reported for the default and the tuned parameter values, we observe that when we parameterize the comparison methods on the development set, they obtain higher scores on the final evaluation corpus. This shows that the optimum values of these parameters are dependent on the context in which we use the summarizers. All the parameterized systems report higher scores using their optimum parameter values, however, only one system (SweSum) reaches a higher rank after parameter tuning. Despite the improvement in the obtained scores, the ranks of other systems do not change compared to their ranks before parameterization.

### 5.3. Limitations

This study has some limitations. First, it seems that four systems of the comparison methods, i.e. SweSum, SUMMA, TexLexAn, and AutoSummarize, do not benefit from common preprocessing procedures, stop-word removal and stemming. This suggests that if they had been provided with the output of such preprocessing steps, their performance could be increased. However, if they had been evaluated with preprocessed documents, the obtained ROUGE scores would not have been indicative of their actual performance. In real applications, the majority of users may give these systems text documents without applying any preprocessing procedure. Second, there may be some similarities between concepts within an input document, and this is not considered by the itemset-based summarizer. Dealing with these similarities may require to gain access to additional resources that contain the similarity information and might increase the complexity of the system.

## 6. Conclusion

In this paper, we propose a novel biomedical text summarization method employing a well-known data mining technique, namely itemset mining, and concept-based modeling of the text. The summarizer maps the input document to the concepts contained in the UMLS to facilitate the concept-level analysis of the source text. With the use of concept-level analysis rather than traditional term-based approaches, the itemset-based summarizer is better equipped to address the inherent ambiguities of biomedical text. The summarizer utilizes frequent itemset mining to discover recurrent and correlated concepts appearing together in the source text. The discovered frequent itemsets demonstrate the important themes of the input document. The itemset-based summarizer quantifies the relatedness and informativeness of each sentence using the support values of the frequent itemsets covering the sentence. Eventually, it produces the final summary by putting the most informative sentences together. We evaluate the itemset-based summarizer on a collection of 400 biomedical scientific articles. Compared with other summarization methods that utilize statistical and similarity features or word frequency algorithms, the itemset-based summarizer shows better summarization performance. The evaluation results confirm that the use of itemset mining in combination with domain knowledge provides an useful method to generate an accurate concept-based model for this type of summarization.

We also investigate the role of the minimum support threshold involved in the itemset mining algorithm on the quality of produced summaries. The support values less than the optimum threshold value generate a large number of itemsets that mislead the summarizer and reduce its accuracy by redundant information. Moreover, the support values greater than the optimum threshold value produce fewer itemsets. In this case, the summarizer's knowledge is not sufficient, and its accuracy decreases due to incomplete information concerning the essential subtopics.

Although the itemset-based summarizer show its efficiency and usefulness for single-document biomedical text summarization, it may encounter the redundancy problem in multi-document summarization. We intend to concentrate on this issue by incorporating a redundancy reduction strategy into the summarization method and proposing a multi-document summarization method. Our future work will also include extending the itemset-based summarization modeling for query-focused biomedical text summarization. As another topic for future research, GraphSum could be extended for biomedical text summarization, dealing with concepts rather than terms.

## 7. Mode of availability

The Java source code of the itemset-based biomedical text summarizer and its documentation are accessible at http://dkr.iut.ac.ir/content/code-itemset-based-summarizer.